\documentclass[twoside]{book}
\usepackage{graphicx}       

 \makeatletter
\ifx\UNDEF\mail\def\mail{ }\else\fi
\ifx\UNDEF\prange\def\prange{0 0}\else\fi

\gdef\@empty{}
\def\Mail#1 #2 {\gdef\thecontact{#1}\gdef\theaddr{#2}}
\def\Range#1 #2 {\gdef\thefirstpage{#1}\gdef\thelastpage{#2}}
{\let\'\mail \expandafter\Mail\' }	
{\let\'\prange \expandafter\Range\' }	
 \gdef\@shtitle{\relax}
 \long\def\shtitle#1{\gdef\@shtitle{#1}}
 \long\def\author#1{\gdef\@author{#1}}
 \def\affil#1{\par\noindent{\rm#1\par}}
 \gdef\@abstract{}
 \long\def\abstract#1{\gdef\@abstract{#1}}

 \renewcommand{\@evenhead}{\thepage\qquad\qquad\@shtitle\hfil}
 \renewcommand{\@oddhead}{\hfil\@shtitle\qquad\qquad\thepage}

 \def\maketitle{\thispagestyle{empty}\chapter{\@title}}
 \renewcommand\chapter{\if@openright\cleardoublepage\else\clearpage\fi
                    \thispagestyle{empty}%
                    \global\@topnum\z@
                    \@afterindentfalse
                    \secdef\@chapter\@schapter}
 \def\@makechapterhead#1{%
  \vspace*{50\p@}%
  {\parindent \z@ \raggedleft \normalfont
    \ifnum \c@secnumdepth >\m@ne
      \if@mainmatter
        \huge \@chapapp{} \thechapter
        \par\nobreak
        \vskip 20\p@
      \fi
    \fi
    \interlinepenalty\@M
    \Huge \bfseries #1\par\nobreak
    \vskip.25in
    \large\bfseries\@author\par\nobreak
    \vskip 40\p@}
    \ifx\@abstract\@empty\else{\small\@abstract\par\vskip20\p@}\fi
  }

\DeclareRobustCommand\em
        {\@nomath\em \ifdim \fontdimen\@ne\font >\z@
                       \upshape \else \slshape \fi}

\def\@begintheorem#1#2{\sl \trivlist \item[\hskip \labelsep{\bf #1\ #2}]}
\def\@opargbegintheorem#1#2#3{\sl \trivlist
     \item[\hskip \labelsep{\bf #1\ #2\ (#3)}]}

 \newcommand{\fig}[1]{Fig.~\ref{fig:#1}}

 \newcommand{\sectlabel}[1]{\label{sect:#1}}
 
 \newcommand{\figlabel}[1]{\label{fig:#1}}

  \def\@arabic#1{\number #1} 

\long\def\@makecaption#1#2{
	\vskip\abovecaptionskip
	\sbox\@tempboxa{{\small {\bf #1}: #2}}%
	\ifdim\wd\@tempboxa>\hsize
	    {\small {\bf #1}: #2\par}
	\else
	   \global\@minipagefalse
	   \hbox to\hsize{\hfil\box\@tempboxa\hfil}
	\fi
	\vskip \belowcaptionskip}

\def\figstrut#1{\hbox to\linewidth{\vrule height#1\hfill}}

\newcommand{\Fig}[4][!htb]{
\begin{figure}[#1]
 \centering\leavevmode#3%
 \caption{#4}
 \figlabel{#2}
\end{figure}                 }

\renewenvironment{thebibliography}[1]
     {\section*{\bibname
        \@mkboth{\MakeUppercase\bibname}{\MakeUppercase\bibname}}%
      \list{\@biblabel{\@arabic\c@enumiv}}%
           {\settowidth\labelwidth{\@biblabel{#1}}%
            \leftmargin\labelwidth
            \advance\leftmargin\labelsep
            \@openbib@code
            \usecounter{enumiv}%
            \let\p@enumiv\@empty
            \renewcommand\theenumiv{\@arabic\c@enumiv}}%
      \sloppy
      \clubpenalty4000
      \@clubpenalty \clubpenalty
      \widowpenalty4000%
      \sfcode`\.\@m}
     {\def\@noitemerr
       {\@latex@warning{Empty `thebibliography' environment}}%
      \endlist}
\makeatother

\shtitle{A two-dimensional rough surface: Experiments on
a pile of rice.}  
\title{A two-dimensional rough surface: Experiments on
a pile of rice.}
\author{C.~M.~Aegerter, R.~G\"unther, and R.~J.~Wijngaarden
\affil{Department of Physics and Astronomy; Vrije Universiteit;
1081HV Amsterdam; The Netherlands.}}

\abstract{Dynamical roughening of interfaces has received much
attention in recent years. However, experiments have been
restricted to one dimensional (1d) systems. Moreover, theoretical
studies of the two dimensional (2d) case have been highly
inconclusive. Here we introduce an experimental 2d system, with
which the theories can be tested. As is shown, the surface of a 2d
pile of rice shows roughening behaviour in both space and time,
with a roughness exponent $\alpha_{2d}$ = 0.39(3) and a growth
exponent $\beta_{2d}$ = 0.27(3). }

\begin{document}
\maketitle
\section{Introduction}\sectlabel{intro}

Roughening phenomena of interfaces have been studied extensively
in recent years due to their wide range of applicability. Rough
interfaces appear in such diverse systems as flux propagation in
superconductors \cite{radu}, the burning of papers \cite{jussi},
diffusion waves \cite{marco}, bacterial colonies \cite{bacteria},
flow through porous media \cite{porous} and many more
\cite{barabasi}. Even though all of these systems have very
different microscopic physics governing the processes, they can be
described by simple models from a very small number of
universality classes \cite{barabasi}. The most famous such model
is described by a non-linear diffusion equation known as the
Kardar-Parisi-Zhang (KPZ) equation \cite{KPZ}:
\begin{equation}
\partial_t h(x,t) = \nu \Delta h(x,t) + \lambda (\nabla h(x,t))^2 + \eta (x,t),
\end{equation}
where $\nu$ is the diffusion coefficient, $\eta$ is a noise term,
$\lambda$ quantifies the non-linearity and $h(x,t)$ is the
position of the interface. In one dimension, the scaling behavior
of an interface governed by the KPZ equation can be analytically
solved. The roughness is parameterized by the width of the
interface given by:
\begin{equation}
\sigma(t,L) = \left(\langle (h(x,t) - \langle h(t) \rangle_L)^2
\rangle_L \right)^{1/2}.
\end{equation}
Here, $\langle \cdot \rangle_L$ denotes the average over the
interface in space. For a self-affine surface, the width growth as
a power law in time $\sigma \sim t^\beta$, until saturation is
reached when the correlation length becomes comparable to the
system size \cite{barabasi}. This growth exponent, $\beta$,
characterizes the dynamics of the process. After the saturation
time, the width is constant in time at a value $\sigma_{sat} (L)
\sim L^\alpha$, which grows as a power law with the system size
\cite{barabasi}. This roughness exponent, $\alpha$, characterizes
the structure of the interface. For the KPZ equation, one obtains
$\alpha$ = 1/2 and $\beta$ = 1/3 \cite{KPZ}.

For a multi-dimensional KPZ system however, the theoretical
situation is unclear. Analytical treatments of the KPZ equation
only exist in approximations \cite{amar,cates,toral}, and results
from numerical simulations vary greatly \cite{barabasi}.
Similarly, experiments have up to now been restricted to a single
dimension. The experimental problem of a two dimensional rough
surface asks for a surface reconstruction technique with enough
spatial resolution to span some orders of magnitude, while at the
same time having the temporal resolution to capture the dynamics
of the process, which is not easily achieved. Secondly, a system
has to be found that exhibits KPZ roughening in 2d and is
accessible experimentally. Presently, there is some interest to
combine the exact results on the KPZ equation with the concept of
self-organized criticality (SOC) \cite{SOC}. The surface of a
sandpile, which is the archetypal system to study SOC, can be
mapped onto a system which follows KPZ dynamics \cite{alava}. This
is intriguing since it brings together two established fields of
research, however has not yet been tested experimentally. We study
here the front of a 2d rice-pile and its roughening behavior,
showing that it does indeed obey KPZ dynamics. We choose rice,
since it has been shown in 1d that a rice-pile does indeed show
SOC behaviour \cite{frette}. With the surface of a rice-pile
established as a roughening system, we can extend the study
further to include the full 2d surface of the pile spanning an
area of $\sim$1x1m$^2$. Using this system, we can determine
roughening and growth exponents in 2d and compare them with
theoretical predictions \cite{amar,cates,toral}.

In section
2, we will discuss the experimental setup including the surface
reconstruction technique based on active-light stereoscopy, as
well as the growing mechanism of the pile. In section 3, we
develop the analysis techniques, with special emphasis on the
generalization from known 1d methods to the 2d problem. In section
4 we present the results of the rice-pile experiment. There, it is
first shown that the front behaviour in 1d does in fact obey KPZ
scaling before discussing the 2d results. Those results are used to
put constraints on theoretical results for 2d KPZ behavior.
\medskip

\section{Experimental setup}\sectlabel{setup}

The rice-pile is grown by dropping rice, uniformly distributed
along a line using a custom-built dispenser. The dispenser
consists of a distribution board and a sowing machine. In the
sowing machine, an eccentric rotor keeps the rice in motion such
that a steady flow of rice is achieved at the rate of $\sim$5 g/s.
This flow of rice is subsequently distributed along a line of 1 m
in the distribution board using simple geometry (see
\fig{dist}). The principle is related to that of a pin-board
producing a Gaussian distribution.

\Fig{dist}{\input{epsf} \epsfxsize 7cm
\centerline{\epsfbox{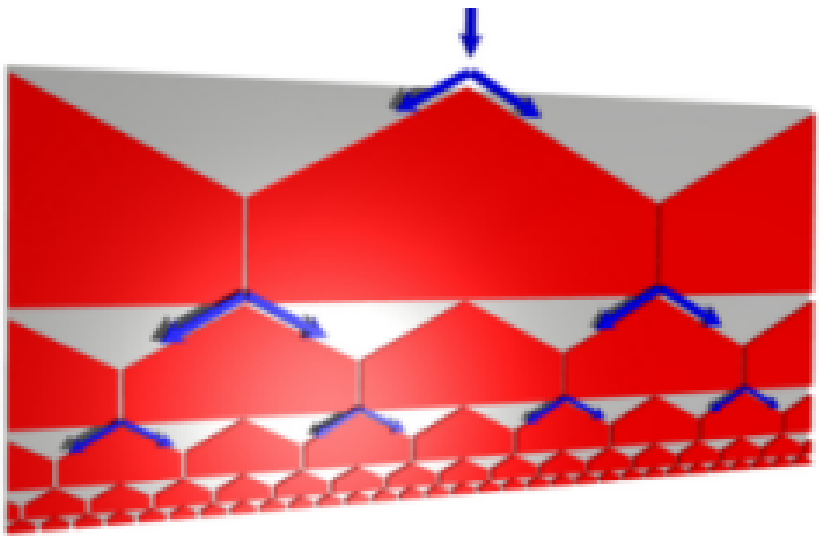}}} {A schematic image of the
distribution board. Rice is dropped from a single point on the top
and subsequently divided into even compartments. At the end a line
of rice uniformly distributed in 64 intervals is obtained, which
is used to grow the rice-pile at a rate of $\sim$5 g/s.}

In order to study the surface properties of the rice-pile, a 3d
reconstruction technique was developed, based on active-light
stereoscopy \cite{rad}. A set of colored lines is projected onto
the pile at approximately right angles using an overhead
projector. In the stereoscopic view \cite{stereo}, the projector
takes the place of the second camera passing its information to
the camera via the colored lines. The camera itself is placed at
an angle of 45 degrees to the surface of the pile and the
projected lines. From this view-point the projected lines are
deformed and can be used to determine the 3d structure of the
surface in the same way as iso-height-lines do on a map. An
example of such a reconstruction is shown in \fig{recons}.
Measurements on test objects show that a surface of 1x1 m$^2$ can
be reconstructed with an accuracy of 1-2 mm, which is comparable
to the size of the rice grains and thus suited for
the present purpose. The use of differently colored lines allows
for better filtering and thus for better identification of the
lines in the computer.

\Fig{recons}{\input{epsf}
\epsfxsize 7cm
\centerline{\epsfbox{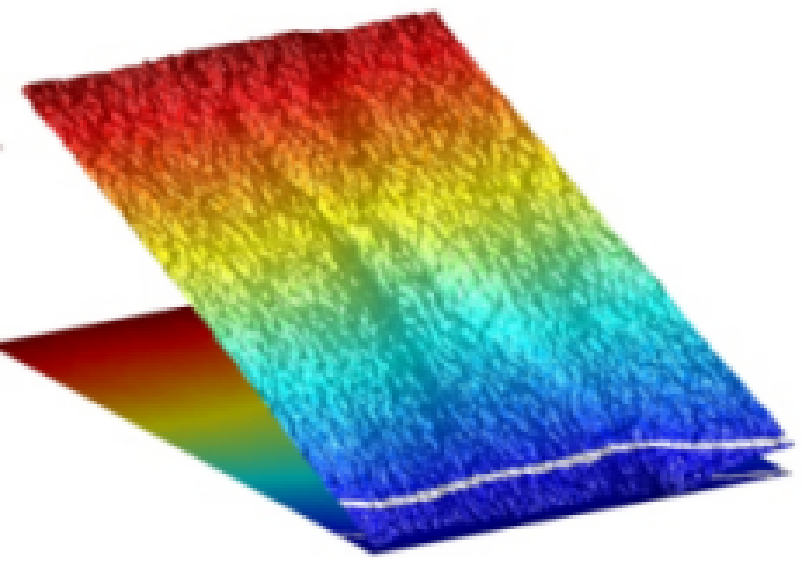}}}
{Reconstruction of the surface of a rice-pile. The white line
indicates the position of the growing front.}

\section{Analysis methods}

As noted in the introduction, rough surfaces are often analyzed
using the width of the interface to characterize its structure and
dynamics. However, in order to obtain reliable results, many
experiments have to be averaged over, using such a method
\cite{barabasi}. A more promising way of analysis, which has been
extensively used in the analysis of 1d experiments is via the
two-point correlation function \cite{barabasi}
\begin{equation}
C(x,t) = \langle (h(\xi,\tau) - h(\xi+x,\tau+t))^2\rangle_{\xi,\tau}^{1/2}.
\end{equation}
In both space and time the scaling behaviour of the correlation
function is the same as that of the width thus making it possible
to determine the growth and roughness exponents from $C(x,t)$. In
addition, the growth exponent can be determined from data obtained
after the saturation time, since in the correlation function only
time differences are important.

When generalizing the method to 2d, computational difficulties arise.
Because of the number of points to compare, the number of operations
to be carried out to determine the correlation function grows with
the fourth power of the size of the surface. However, tests on small
surfaces indicate that the radial average of $C(x,y,t)$ scales like the
2d local width, but due to the computational inefficiency we were
using yet another method to determine the roughness and growth exponents.

The power spectrum, or structure function \cite{power}, can be
determined easily for 1d and 2d systems from the square of the
Fourier transform $\hat{h}(k_x,k_y)$ of the local height $h(x,y)$
\begin{equation}
S(k_x,k_y) = |\hat{h}(k_x,k_y)|^2.
\end{equation}
Here, the computational load is just given by determining the 
Fourier transforms, which also in 2d only grows with the square 
of the size of the surface thus making it
feasible to calculate the distribution function of the whole
rice-pile surface. The square root of the integral of $S(k_x,k_y)$
over k-space, the distribution function $\sigma(k_x,k_y)$,
\begin{equation}
\sigma(k_x,k_y) = \left(\int\int S(k_x,k_y) dk_xdk_y \right)^{1/2}
\end{equation}
is equal to the rms-width of the interface \cite{power}. Thus the
distribution function also has the same scaling behaviour as the
width and can therefore be used to determine the roughness and
growth exponents. Again, a radial average of $\sigma(k_x,k_y)$,
$\sigma(k)$, scales like the 2d local width, thus making
comparisons with previous simulations of ballistic deposition
models possible. Moreover, real 2d measures like the distribution function
can also give information about anisotropies of the scaling in the
x- and y-directions.

The distribution function is also useful in investigations of the dynamics
of the processes. In that case, the square of the Fourier transform
$\hat{h}(\omega)$ of the time dependence $h(t)$ has to be determined. The
Fourier transforms are determined using an FFT algorithm after padding the
data with zeros to the next power of two.

\section{Results and discussion}

In order to determine that the rice-pile surface does in fact
follow KPZ behaviour we first determined the roughness and growth
exponents of the front of the pile, 
\Fig{1d}{\input{epsf}
\epsfxsize 10cm
\centerline{\epsfbox{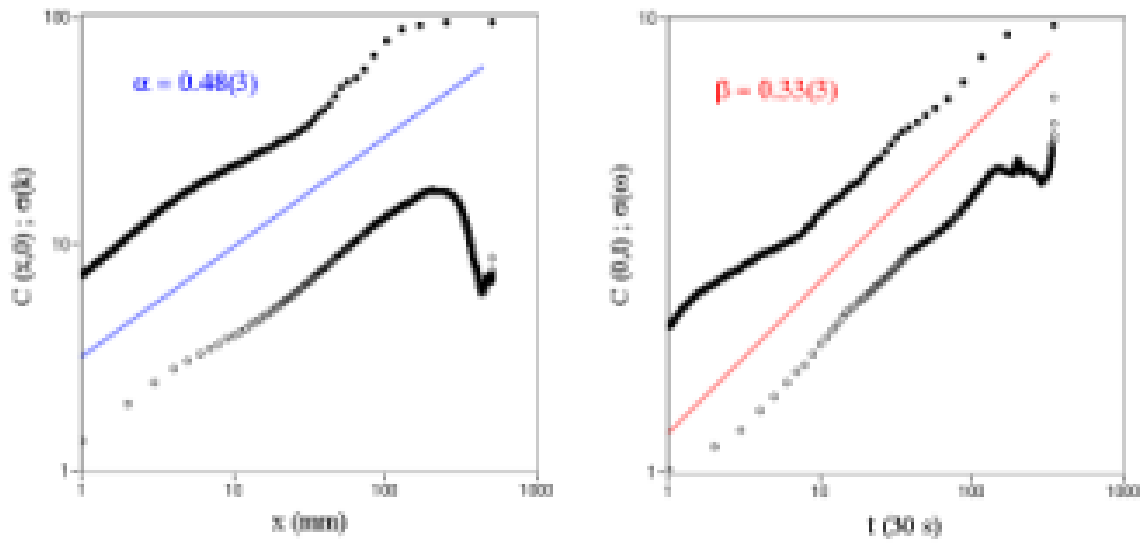}}}
{The behavior of the front of a propagating rice-pile. Both the
correlation function (open symbols) and the distribution function
(full symbols) show scaling in space and time over two decades.
The resulting roughness and growth exponents are in excellent
agreement with the KPZ universality class.}
\noindent given by the line of equal
height of the pile at 0.1 m. The distribution functions determined
in both space, $\sigma(k)$, and time, $\sigma(\omega)$, as well as
the correlation 
functions $C(x,t)$ can be seen in \fig{1d}, where
the values of $\alpha = 0.48(3)$ and $\beta = 0.33(3)$ can be
inferred. These values are in excellent agreement with the
expectations from the KPZ equation thus establishing that KPZ
behaviour does appear in SOC systems.

The 2d distribution function, $\sigma(k_x,k_y)$, which
characterizes the roughening of the whole surface is shown in
\fig{2dsigma} on a triple-logarithmic plot. In the insert, the
angular dependence of a power-law fit to $\sigma$ is shown. This
indicates a dependence of the roughness exponent $\alpha$ on the
direction, which shows the anisotropy of the system. Such an
anisotropy most probably arises from the growth mechanism of the
pile, which is seeded from a horizontal line, thus breaking the
symmetry of the x- and y-directions. It should be noted here that
the exponents corresponding to the x- and y-directions do not have
to agree with those determined in a 1d analysis. This is because
$\sigma(k_x,0)$ already includes data from the y-direction due to
the complex nature of the Fourier transform. Thus, $\sigma(k_x,0)$ 
already presents an effective 2d measure.

\Fig{2dsigma}{\input{epsf}
\epsfxsize 9cm
\centerline{\epsfbox{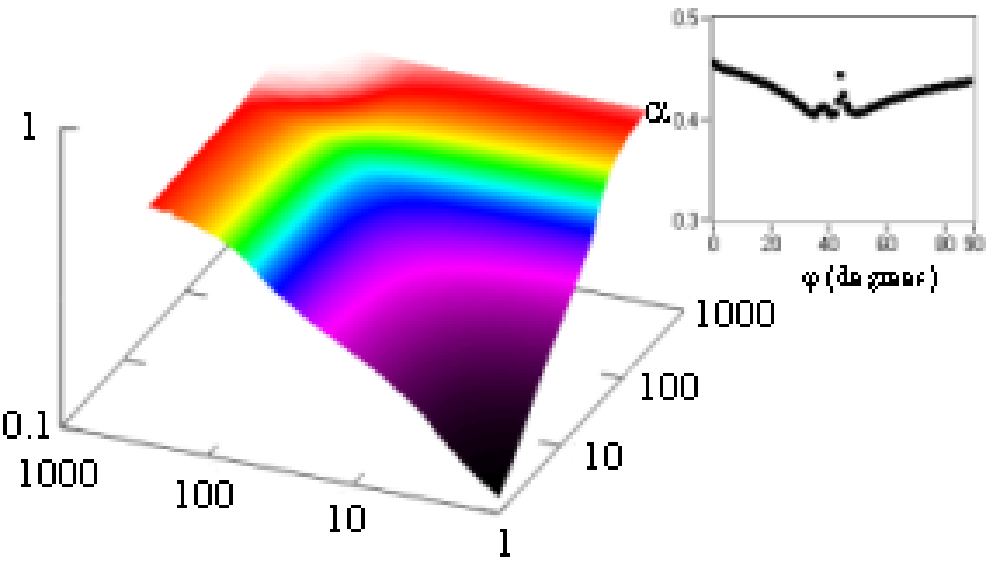}}}
{The 2d distribution function for the rice-pile surface on a triple-logarithmic
plot. From a radial average, the roughness exponent can be determined.
The insert shows an angular dependence of the roughness exponent,
with an anisotropy in the x- and y-directions.}

The radial average of the distribution function, $\sigma(k)$, is
shown in \fig{radial}a, where the value of the roughness exponent
can also be determined. We obtain $\alpha_{2d} = 0.39(3)$, which
is also in agreement with the average of the exponents determined
as a function of angle. In addition, the temporal behavior,
$\sigma(\omega)$ is shown in \fig{radial}b, where we determine the
2d growth exponent to be $\beta_{2d} = 0.27(3)$. Both the
roughness and growth exponents determined experimentally are in
very good agreement with the conjecture derived from
solid-on-solid models by Kim and Kosterlitz \cite{kim} for higher
dimensional exponents given by $\alpha$ = 2/(d+3) and $\beta$ =
1/(d+2). Numerical results from integrating the 2d KPZ equation
vary greatly, with values of $\alpha_{2d}$ ranging from 0.18
\cite{toral} to 0.39 \cite{amar} and $\beta_{2d}$ ranging from 0.1
\cite{toral} to 0.25 \cite{amar}. Our experimental results are in
good agreement with the numerical values of Amar and Family
\cite{amar}, as well as Bouchaud and Cates \cite{cates}
corresponding to the high range of the values, while excluding
most of the other numerical investigations into 2d KPZ behavior.

\Fig{radial}{\input{epsf}
\epsfxsize 10cm
\centerline{\epsfbox{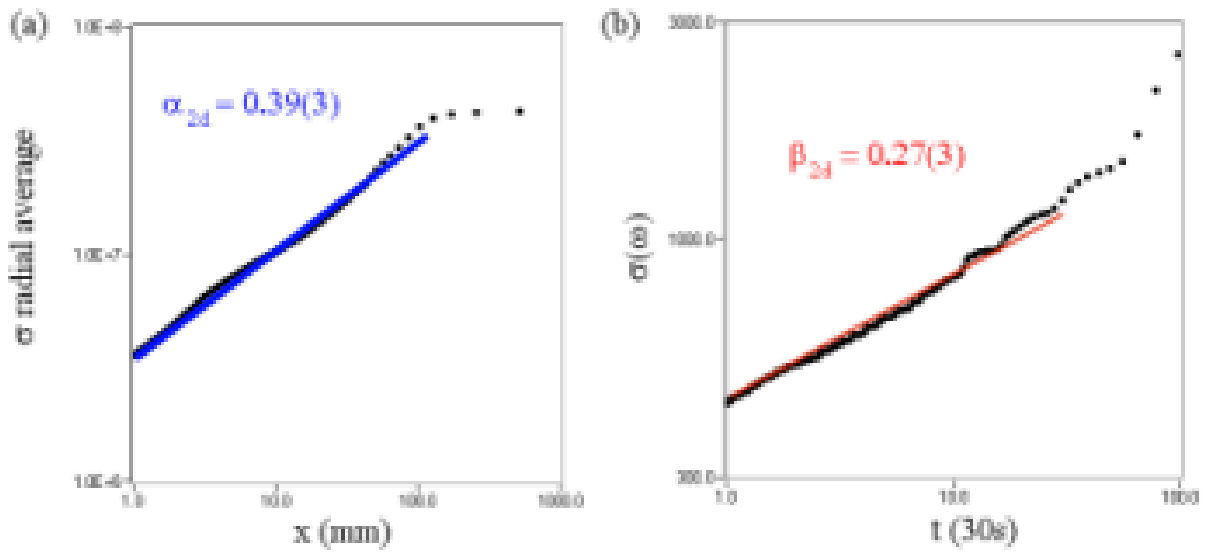}}}
{(a) The radial average of the 2d distribution function, allowing the
determination of the roughness exponent in 2d to be $\alpha_{2d}$ =
0.39(3) form a scaling regime spanning a decade and a half. (b)
The distribution function in time, leading to a growth exponent of $\beta_{2d}$
= 0.27(3). }

\section{Conclusions}

We have presented an experimental study on roughening in a 2d
system. The surface of a rice-pile is measured with a
reconstruction technique based on active-light stereoscopy. In 1d,
the fronts of the rice as the pile is grown shows excellent
agreement with the 1d KPZ universality class with exponents
$\alpha$ = 0.48(3) and $\beta$ = 0.33(3) from a scaling-regime
spanning more than two decades. Thus having established the KPZ
nature of the system under study, we analyze the full 2d surface
of the pile, where find a roughness exponent of $\alpha_{2d}$ =
0.39(3) and $\beta_{2d}$ = 0.27(3). This is consistent with
numerical simulations for ballistic deposition models
\cite{kessler, meakin} and puts a strong experimental constraint
on the available results on 2d KPZ simulations. Our results are in
good agreement though with the results of Amar and Family
\cite{amar}, as well Bochaud and Cates \cite{cates} from numerical
integration of the 2d KPZ equation. In addition however, we have
studied the dependence of the exponent on the direction, where we
find that the system is anisotropic with a somewhat higher
exponent along the front direction. This is probably related to
the difference between the two directions due to the growth
mechanism.

\section{Acknowledgements}

This work was supported by the Swiss National Science Foundation 
and by FOM (Stiching voor Fundamenteel
Onderzoek der Materie), which is financially supported by NWO
(Nederlandse Organisatie voor Wetenschappelijk Onderzoek).


\end{document}